\documentclass[10pt,reqno]{amsart}
\usepackage[utf8]{inputenc}
\usepackage{graphicx}
\usepackage{hyperref}
\usepackage[left=1in, right=1in, bottom=1in,top=1in]{geometry}

\usepackage{amsmath}
\usepackage{amsfonts}
\usepackage{amsthm}
\usepackage{amssymb}

\newtheorem{theorem}{Theorem}

\newcommand{\E}[2][]{\mathbb{E}_{#1}\left[#2\right]}
\newcommand{\Prb}[2][]{\mathbb{P}_{#1}\left(#2\right)}
\newcommand{\Var}[2][]{\mathrm{Var}_{#1}\left(#2\right)}

\newcommand{\Cov}[2][]{\mathrm{Cov}_{#1}\left(#2\right)}
\newcommand{\eq}[1]{\textbf{Equation #1}}
\newcommand{\thm}[1]{\textbf{Theorem #1}}

\newcommand{\bone}[0]{\mathbb{I}}
\newcommand{\tref}[0]{t_{\mathrm{ref}}}

\usepackage{tikz}
\usetikzlibrary{shapes, arrows, positioning,decorations.text}
\tikzset{
    rbox/.style = {rectangle, draw, rounded corners, minimum height=2em, align=center},
    ebox/.style = {rectangle, draw, minimum height=2em, align=center},
    connect/.style ={draw, -latex, thick}
}

\title{A genealogical interpretation of fixed and random effects models of complex traits}

\usepackage{amsaddr}
\author{Hanbin Lee$^{\dagger,\ddagger}$}
\address{
    $^{\dagger}$Department of Medicine, Seoul National University College of Medicine\\Seoul, Republic of Korea\\
    $^{\ddagger}$Department of Mathematical Sciences, Seoul National University\\Seoul, Republic of Korea
    }
\email{hanbin973@snu.ac.kr}

\begin{document}

\maketitle

\begin{abstract}
    In the analysis of complex traits, genetic effects are frequently modelled as either fixed or random effects.
    Such assumptions serve as a foundation of defining heritability and relatedness using genome-wide single nucleotide polymorphism (SNP) markers.
    In this work, I propose a genealogical framework connecting the two assumptions conditional on the ancestral recombination graph (ARG).
    It turns out that the reference time point in which the probability is defined determines whether the effect of a variant to behave as either a fixed or random effect.
    This lays a connection between the PC regression and linear mixed model (LMM) for genetic association study.
    The framework induces a genetic relatedness matrix (GRM) in which the elements are a function of the time to the recent common ancestor.
    Subsequently, a novel trait variance decomposition respect to the regions sharing a common genealogy is followed.
    The variance decomposition then provides a natural means to define heritability and a potential method for gene mapping.
\end{abstract}

\section{Introduction}

The polygenic model of complex traits models the trait as a linearly additive function of genetic variants.
The model dates back to Fisher and played a pivotal role in combining Mendelianism and Darwinism \cite{Fisher_1919}.
The model is now adopted in a wide range of applications including heritability estimation \cite{Yang_2010}, genome-wide associations study (GWAS) \cite{Price_2006, Kang_2010, Zhou_2012} and summary statistics based methods \cite{Bulik_Sullivan_2015, Shi_2016}, to name a few.

Although the functional form of the polygenic model is shared across studies, additional modelling assumptions are frequently made.
Among such assumptions, random effects models of genetic effects is commonly adopted.
The genetic random effects model was first introduced in breeding literature by Henderson \cite{Henderson_1974}, and was later introduced to human genetics \cite{Yang_2010}.
Linear mixed model (LMM), which includes both fixed and random effects, was independently adopted to gene mapping in model organisms, and subsequently in humans \cite{Kang_2008, Kang_2010}.

The unique feature of genetic random effects is that the randomness is for the features (or the columns) and not the samples (or the rows).
Most applications of random effects outside genetics employ random effects to model the row-wise variability, for example, nested samples within groups \cite{Gelman_2006}.
In such a case, the source of randomness can be attributed to sampling variation.
However, it cannot explain the column-wise random effects in genetic applications.

In this work, I show that the column-wise variability of genetic random effects can be attributed to randomly occurring mutations in the genome.
For a given reference time point in the past \cite{Weir_1984, Thompson_2013}, polymorphic loci appears randomly through mutations which can be thought as sampling trait-associated sites randomly.
This serves as the basis of column-wise randomness of genetic random effects.
In turn, mutations that have emerged prior to the reference time point behave as fixed effects.
This observation suggests a putative connection between PC regression and LMM used in genetic association studies.

The resulting genetic relatedness matrix (GRM) closely resembles the version of McVean \cite{McVean_2009} which was first developed to interpret principal component analysis (PCA) applied to the genotype matrix.
Under neutral evolution, the elements of the GRM is a function of pairwise coalescence time which is a natural measure of relatedness.
The trait variance can be decomposed into local regions sharing the same genealogical history that gives a natural way of defining local heritability and a potential method to perform quantitative trait loci (QTL) mapping.

\section{Results}

\subsection{The ancestral recombination graph}

The ancestral recombination graph (ARG) is a collection of evolving local coalescent trees as we navigate through the genome \cite{Futuyma1992-bj, GRIFFITHS_1996}.
The tree at a fixed position is called the marginal tree.
Nearby trees are correlated due to recombination events.
Let $l=1,\ldots,n_l$ be the index of loci across the genome. 
$n_l$ is the number of regions with different marginal tree topology.
Sites indexed by $s$ are nested within a locus ($s \in l$) and we denote the number of sites within locus $l$ as $n^l$.

The marginal tree at locus $l$ is denoted as $T^l$.
We index the branches of a tree as $b$ and write its length as $L(b)$.
The total length of $T^l$ is $L^l = \sum_{b \in T^l} L(b)$.
We override the the notation to write $b \in T^l$ which means the the branch is within $T^l$.
The mutation rate of a locus is denoted as $u^l$.
As there are $n^l$ sites within the locus, the per-site mutation rate is given by $u^l/n^l$.
The number of total mutations on $T^l$ follows a Poisson distribution with mean $u^l \sum_{b \in T^l} L(b) = u^l L^l$.
This implicitly assumes that the mutation is neutral so that the mutation does not alter the topology of the tree.

We make the infinite sites assumption so that every time a mutation occurs, it appears at a new site \cite{Kimura_1969}.
Therefore, back mutation is ignored.
In this work, the ARG is given fixed and mutations occur at random.
The general theory of this setting has been studied by Ralph \cite{Ralph_2019}.

\subsection{An interpretation of the genetic random effects}

Let $i$ be the index of haploids and $s$ be the index of sites.
We write the row indices as subscripts and the column indices as superscripts.

Let $\bone^s$ be the indicator function of site $s$ being polymorphic.
The total number of sites is denoted as $N_s$.
Assuming the infinite sites model, a mutation occurs only once at a site with probability proportional to the total length of the tree at the locus \cite{Kimura_1969}.
$G_i^s=0,1$ is the genetic value of individual $i$ at site $s$.
Then the polygenic model can be written as
\begin{align}
    y_i = \sum_l \sum_{s \in l} G_i^s\beta^s \bone^s + \beta^0 + \epsilon_i
    \label{eq:polygenic_ind}
\end{align}
where $\beta^0$ is the intercept and $\epsilon_i$ is the random error.

We consider $n$ samples.
The trait vector, the genotype matrix, the error vector of the sample are $\mathbf{y} = (y_1, \ldots, y_n)^T$, $\mathbf{G}^s = (G_1^s, \ldots, G_n^s)^T$ and $\boldsymbol{\epsilon}=(\epsilon_1, \ldots, \epsilon_n)^T$.
\eq{\ref{eq:polygenic_ind}} can be written as 
\begin{align}
    \mathbf{y} = \sum_l \sum_{s \in l} \mathbf{G}^s \beta^s \bone^s + \beta^0\mathbf{1}_{n} + \boldsymbol{\epsilon}
    \label{eq:polygenic_sample}
\end{align}
where $\mathbf{1}_n = (\overbrace{1, \ldots, 1}^{n\text{ copies}}$$)^T$.

Here, each $\beta^s$ are assumed to be constant as the variant at a fixed position in the genome is likely to have the same effect.
Instead, $\bone^s$ and $G_i^s$ are random variables respect to the mutational process emerging on the ARG.
Each evolutionary trial beginning at a reference time point from the past, we observe a different set of variants corresponding to their effect size because mutations occur at random positions.
The distribution function of this random variable $\beta$ is therefore
\begin{align}
    \begin{aligned}
        \Prb{ \beta \leq x} &= \sum_{s:\beta^s\leq x} \Prb{\bone^s=1} \\
        &= \sum_l \sum_{s:\beta^s \leq x, s\in l} \frac{u^lL^l}{n^l} e^{-u^lL^l/n^l} \\ 
        &\approx \sum_l u^lL^l \cdot \frac{\#\{\text{sites $s$ such that $\beta^s \leq x$}\}}{n^l}
    \end{aligned}
    \label{eq:gen_arch}
\end{align}

The effect size distribution is often termed as the genetic architecture \cite{Timpson_2017}.
According to \eq{\ref{eq:gen_arch}}, the genetic architecture of a trait depends on the mutation rate and the topology of the ARG.

\subsection{The reference time point}
Let the current time $t_{0} =0$ and the time to the most recent common ancestor of the population as $-t_{\mathrm{MRCA}} <0$.
$\tref \in [-t_{\mathrm{MRCA}}, t_0]$ is the reference time point.
Time to MRCA can be different across loci.
We add a superscript to denote this difference $t_{\mathrm{MRCA}}^l$.
The value is constant within a locus sharing the local tree.

At $t_{\mathrm{MRCA}}$, all sites are monomorphic so $\bone^s=0$ for all $s$.
Since mutations accumulate over time, the number of polymorphic sites $n_s$ increases with time.
Therefore, $\bone^s$ increases over time, and is a monotone stochastic process on $[-t_{\mathrm{MRCA}}, t_0]$.
I denote the associated filteration as $\mathcal{F}(t)$.
Then the conditional expectation is defined as $\E[t]{\;\cdot\;}:= \E[]{\;\cdot \mid \mathcal{F}(t)}$. 
Once the mutation has arrived at time $t$, $\bone^s$ ($s=1,\ldots,N_s$) and $G_i^s$ ($i=1,\ldots,N$) remain constant.

\subsection{The genetic relatedness matrix and the variance decomposition}

We now assume that $G_i^s$ is independent from $\epsilon_i$.
Since $G_i^s$ is determined by the mutational process holding the ARG fixed, this means the mutational process is independent from $\epsilon_i$.
Let $S(t)$ be the set of sites in which mutations have arrived before time $t$.
Also, $h_t^l=\mathrm{min}\{t_{\mathrm{MRCA}}^l, t\}$ and $h_{t,ij}^l=\mathrm{min}\{t_{\mathrm{MRCA},ij}^l, t\}$ where $t_{\mathrm{MRCA},ij}^l$ is the coalescence time of individual $i$ and $j$ at locus $l$.
\begin{theorem}
    The trait variance respect to the reference time point $\tref$ is
    \begin{align*}
        \Var[\tref]{\mathbf{y}} = \sum_l (\sigma_{\tref}^l)^2 \mathbf{A}_{\tref}^l + \sigma_e^2 \mathbf{I}_{n \times n}
    \end{align*}
    where
    \begin{gather*}
        \begin{gathered}
            \left(\sigma_t^l\right)^2 = \frac{u^l}{n_l} \sum_{s\notin S(t)\& s\in l} \left(\beta^s \right)^2 \\
            \left[\mathbf{A}_t^l\right]_{ij} = h_t^l - h_{t,ij}^l
        \end{gathered}
    \end{gather*}
\label{thm:var_decomp}
\end{theorem}

Note that the diagonal elements of $\mathbf{A}_t^l$ are $h_t^l-h_{t,ij}^l = h_t^l$ because $h_{t,ij}^l=0$ for $i=j$.
Therefore, one might choose to normalize the matrix with $h_t^l$ to write as
\begin{gather*}
    \begin{gathered}
        \left(\widetilde{\sigma}_t^l\right)^2 = \frac{h_t^lu^l}{n_l} \sum_{s\notin S(t)\& s\in l} \left(\beta^s \right)^2 \\
        \left[\widetilde{\mathbf{A}}_t^l\right]_{ij} = \frac{h_t^l - h_{t,ij}^l}{h_t^l}
    \end{gathered}
\end{gather*}
Then the overall and the local heritability can be defined as
\begin{gather}
    \begin{gathered}
        h_t^2 = \frac{\sum_m \left(\widetilde{\sigma}_{\tref}^m\right)^2}{\sum_m \left(\widetilde{\sigma}_{\tref}^m\right)^2 + \sigma_e^2} \\
        \left(h_t^l\right)^2 = \frac{ \left(\widetilde{\sigma}_{\tref}^l\right)^2}{\sum_m \left(\widetilde{\sigma}_{\tref}^m\right)^2 + \sigma_e^2}
    \end{gathered}
    \label{eq:h2}
\end{gather}
because the diagonal elements are set to $1$.

\thm{\ref{thm:var_decomp}} gives a variance decomposition respect to genetic regions $l$ sharing the same tree topology $T^l$.
As $\sigma_t^l$ is proportional to the sum of squares of $\beta^s$ within the region, a non-zero value indicates a QTL.
$\sigma_t^l$ does not depend on the topology of the tree.
Therefore, the value is the same to the one that would have been obtained from the whole population.
However, the normalized version $\widetilde{\sigma}_t^l$ depends on $h_t^l$ which depends on the tree topology of the sample.
Therefore, the heritability in \eq{\ref{eq:h2}} should be understood as a sample-specific quantity.

\subsection{The fixed-random transition of genetic effects}

Let $S(t)$ be the set of sites where mutations have arrived at time $t$.
After the arrival of the mutation, $\bone^s$ and $G_i^s$ all behave as constants.
$\bone^s=1$ and $G_i^s$ will take either $0$ or $1$ based on the position of the mutation on the local tree.
\begin{theorem}
    The trait mean respect to the reference time point $\tref$ is
    \begin{align*}
        \E[\tref]{y_i} = \sum_l \sum_{s \in l, S(\tref)} \beta^s G_i^s + \sum_l \left( \frac{u^l h_{\tref}^l}{n^l} \cdot \sum_{s \in l, s\notin S(\tref)} \beta^s\right) + \beta^0
    \end{align*}
    which also can be written in vector notation as
    \begin{align*}
         \E[\tref]{\mathbf{y}} = \sum_l \sum_{s \in l, S(\tref)} \beta^s \mathbf{G}^s + \sum_l \left( \frac{u^l h_{\tref}^l}{n^l} \cdot \sum_{s \in l, s\notin S(\tref)} \beta^s\right) \mathbf{1}_n + \beta^0 \mathbf{1}_n
    \end{align*}
    \label{thm:exp_decomp}
\end{theorem}

Looking at \thm{\ref{thm:var_decomp}} and \thm{\ref{thm:exp_decomp}}, at $\tref$, we can see that the time when the site became polymorphic determines whether the genetic effect of the variant at the site behaves as either a fixed or a random effect.
As one moves $\tref$ from the past to the present, genetic effects evolve gradually from random effects to fixed effects.

Sites that have become polymorphic prior to $\tref$ appears in the expectation as seen in \thm{\ref{thm:exp_decomp}}.
However, it does not contribute to the pairwise relatedness in the GRM as shown in \thm{\ref{thm:var_decomp}}.
The exact opposite holds for variants that are yet polymorphic at time $\tref$.
Although these variants do appear in the second term of \thm{\ref{thm:exp_decomp}}, the value is constant for all individuals so it has no discernible effect.

This result reveals a putative connection between PC regression and LMM.
PC regression can be thought as a penalized regression for high-dimensional problems \cite{Hastie_2009}.
The regression equation including all sites in $S(\tref)$ fulfills the condition.
In this perspective, PC regression is a penalized approximation of the regression in \thm{\ref{thm:exp_decomp}}.
Therefore, a pure fixed-effects PC regression can be viewed as a special case at $\tref=t_0$ and the pure random-effects LMM can be viewed as a special case at $\tref=\max_{l=1,\ldots,n_l}\{t_{\mathrm{MRCA}}^l\}$, the grand TMCRA.

\subsection{Quantitative trait loci (QTL) mapping}

The variance decomposition of \thm{\ref{thm:var_decomp}} gives a natural mean to perform quantitative trait loci (QTL) mapping with arbitrary range up to the resolution given by the ARG.
As the true ARG is unknown, the working method is based on the inferred ARG \cite{Speidel_2019,Kelleher_2019,Wohns_2022}.

The likelihood estimator we fit is based on
\begin{align}
    \mathbf{y} \sim \mathcal{N} \left(\sum_l \sum_{s \in l, S(\tref)} \beta^s \mathbf{G}^s + \beta_{\tref}^0 \mathbf{1}_n,  \sum_l (\sigma_{\tref}^l)^2 \mathbf{A}_{\tref}^l + \sigma_e^2 \mathbf{I}_{n \times n} \right)
    \label{eq:likelihood_raw}
\end{align}
where $\beta_t^0 = \sum_l \left( \frac{u^l h_t^l}{n^l} \cdot \sum_{s \in l, s\notin S(t)} \beta^s\right) + \mathbf{1}_n$.
Setting $\tref$ to the grand TMRCA $\max_{l=1,\ldots,n_l}\{t_{\mathrm{MRCA}}^l\}$ will measure the effect of all variants that emerged since the grand TMRCA, and moving $\tref$ closer to the present will capture more recent variants.
Let $\mathbf{V}_n = \sum_l (\sigma_{\tref}^l)^2 \mathbf{A}_{\tref}^l + \sigma_e^2 \mathbf{I}_{n \times n}$.
Then the minus log-likelihood according to \eq{\ref{eq:likelihood_raw}} is
\begin{align}
    \frac{1}{2} \log \vert \mathbf{V}_n \vert +\frac{1}{2} \left(\mathbf{y}-\sum_l \sum_{s \in l, S(\tref)} \beta^s \mathbf{G}^s - \beta_{\tref}^0 \mathbf{1}_n\right)^T \mathbf{V}_n^{-1} \left(\mathbf{y}-\sum_l \sum_{s \in l, S(\tref)} \beta^s \mathbf{G}^s - \beta_{\tref}^0 \mathbf{1}_n\right)
\end{align}
The goal is to find the parameter minimizing the above objective function.
A non-zero $\sigma_{\tref}^l$ can be thought as an evidence for a genetic signal at region $l$.

A critical drawback of estimating \eq{\ref{eq:likelihood_raw}} from real data is that the number of loci is far larger than the number of samples available.
A penalization would be required to properly estimate the parameters \cite{Tibshirani_1996, Hastie_2009}.
Sparse penalization would have an additional advantage of loci selection.

\section{Discussion}

This work presents a genealogical interpretation of the genetic fixed and random effects model.
Under the neutral evolution assumption, the sample variance can be expressed in terms of local GRMs and the associated variances.
This gives a natural way to define time-specific genealogy-based heritability and a mean to detect QTLs using genealogies.
Within the framework, genetic effects transform from random to fixed effects as the reference time point moves closer to the present.

In this work, the ARG is given fixed and mutations occur uniformly on the branches of the ARG following a Poisson process \cite{Ralph_2019}.
As a result, genetic variation is dominated by the mutational process.
The remaining variability comes from the non-genetic factor which was assumed to be independent from the mutational process.
To place mutations uniformly on a fixed ARG, it requires the evolutionary process to be neutral.
Since non-neutral mutations subject to selection alter the topology of the tree, the distribution of mutations depends on the position at the tree under selection.
This will subsequently change the GRM derived in this work.

Similar GRMs have been studied to date. 
The GRM of this study resembles that of McVean in a sense that the genotypes were not normalized \cite{McVean_2009}.
The core difference is that the McVean's GRM allows the genealogy to vary.
As a result, the elements appear as a function of expected branch lengths rather than the branch lengths themselves.
My GRM conditions on the ARG so the branch lengths are given constant which does not require taking expectations.
Other approaches have conditioned on ARG resulting GRMs as a direct function of branch lengths rather then their expectations \cite{Wang_2021, Zhang_2021, Fan_2022, Link_2023}.
The difference is that genotypes are normalized with an associated factor of $\alpha$ in these works \cite{Speed_2014}.
My GRM is identical to the $\alpha=0$ case up to a multiplicative constant.

This work suggests a novel form of narrow sense heritability.
The advantage of the novel definition is that it naturally breaks down into local regions without making assumptions about the linkage disequilibrium (LD).
As mutation at different sites are assumed to be independent, conditioning the ARG removes the correlation between sites.
Therefore, LD can be ignored which gives local genetic variances that simply adds up to the overall genetic variance.
This is in contrast with previously proposed SNP-based heritability where the total genetic variance is not an additive function of local variances due to LD \cite{Yang_2010, Bulik_Sullivan_2015, Shi_2016}.

Although the paper proposes a potential method for gene mapping, the model requires a method to estimate variance components of a high-dimensional linear model making empirical implementation difficult.
Penalized variance component methods have been proposed but does not scale to biobank-scale data \cite{Schaid_2020, Kim_2021}.
An alternative strategy would be testing local GRMs marginally as proposed by Link and colleagues \cite{Link_2023}.
However, the current theory provides very little information on the consequence of testing local GRMs separately rather than as a whole as suggested by the theory.
Further work is warranted to understand the behavior of marginal testing within the framework.

One notable contribution is that it hints a connection between PC regression and LMM.
Existing theoretical analyses have largely focused on the statistical perspectives such as the number of PCs being included \cite{Hoffman_2013, Zhang_2014}.
This work provides an orthogonal viewpoint based on population genetics.
PC regression accounts genetic effects through fixed effects while LMM does so by random effects.
Under my framework, a genetic effect is either fixed or random given $\tref$ so PC regression and LMM can be thought as two extremes at the same continuum.
This might explain hybrid methods that includes both PC and GRMs \cite{Loh_2015, Jiang_2019}.

This work has several limitations.
First, the theory is based on the neutral evolution assumption.
As background and balancing selection have been found to be pervasive, further work is required to incorporate more realistic evolutionary scenarios \cite{Zeng_2018,O_Connor_2019, Simons_2022, Veller_2023}.
Second, a realistic algorithm is required to estimate the heritability decomposition proposed by the theory.
Third, the GRM proposed by the theory does not exactly match the ones that are used in practice.
Therefore, the analysis of previous methods in the light of the current theory should not be taken as decisive.
Finally, the model considered is purely additive so non-linear effects such as gene-environment interaction are ignored.

\appendix
\setcounter{equation}{0}
\renewcommand{\theequation}{A\arabic{equation}}
\section*{Appendix: Proofs}
The following proofs are based on the quantitative trait model and I restate the model for convenience.
\begin{align}
    \mathbf{y} = \sum_l \sum_{s \in l} \mathbf{G}^s \beta^s \bone^s + \beta^0\mathbf{1}_{n} + \boldsymbol{\epsilon}
\end{align}

\begin{proof}[\proofname\ of \textbf{Theorem 1}]

First, expand the variance.
\begin{align}
    \begin{aligned}
        \Var[t]{\mathbf{y}} &= \Var[t]{\sum_l \sum_{s \in l} \mathbf{G}^s \beta^s \bone^s + \beta^0\mathbf{1}_{n} + \boldsymbol{\epsilon}} \\
        &= \sum_l \sum_{s \in l} \Var[t]{\mathbf{G}^s \beta^s \bone^s}+ \Var[t]{\boldsymbol{\epsilon}} \\
        &= \sum_l \sum_{s \in l} \left(\beta^s\right)^2 \Var[t]{\mathbf{G}^s  \bone^s}+ \Var[t]{\boldsymbol{\epsilon}} \\
    \end{aligned}
    \label{proof:var_expand}
\end{align}
The second line follows from the independence of the mutational process across sites.

Next, we compute each element of $\Var[t]{\mathbf{G}^s  \bone^s}$.
\begin{align}
    \begin{aligned}
        \Cov[t]{G_i^s\bone^s, G_j^s\bone^s} &= \E[t]{G_i^s G_j^s \bone^s} -\E[t]{G_i^s \bone^s}\E[t]{G_j^s \bone^s} \\
        &= \frac{h_t^l - h_{t,ij}^s}{L^l} \E[t]{\bone^s} - \frac{h_t^l }{L^l} \E[t]{\bone^s} \cdot \frac{h_t^l}{L^l} \E[t]{\bone^s} \\
        &= \frac{h_t^l -h_{t,ij}^s}{L^l} \cdot \frac{u^l}{n^l} L^l e^{-u^lL^l/n^l} - \frac{h_t^l }{L^l} \cdot \frac{u^l}{n^l} L^l e^{-u^lL^l/n^l} \cdot \frac{h_t^l }{L^l} \cdot \frac{u^l}{n^l} L^l e^{-u^lL^l/n^l} \\
        &\approx (h_t^l -h_{t,ij}^s) \cdot \frac{u^l}{n^l} - h_t^l \cdot \frac{u^l}{n^l} \cdot h_t^l\cdot \frac{u^l}{n^l} \\
        &\approx (h_t^l -h_{t,ij}^s) \cdot \frac{u^l}{n^l} \quad (\because\mbox{$u^l/n^l$ is small}) \\
        &= (h_t^l -h_{t,ij}^l) \cdot \frac{u^l}{n^l} \quad (\because\mbox{the topology is same within $l$})
    \end{aligned} 
    \label{proof:cov_ij}
\end{align}
Substituting \eq{\ref{proof:cov_ij}} to \eq{\ref{proof:var_expand}} gives the desired result.

\end{proof}

\begin{proof}[\proofname\ of \textbf{Theorem 2}]

First, expand the expectation.
\begin{align}
    \begin{aligned}
        \E[t]{\mathbf{y}} &= \E[t]{\sum_l \sum_{s \in l} \mathbf{G}^s \beta^s \bone^s + \beta^0\mathbf{1}_{n} + \boldsymbol{\epsilon}} \\
        &= \sum_l \sum_{s \in l} \E[t]{\mathbf{G}^s \beta^s \bone^s}  + \beta^0\mathbf{1}_{n} \\
        &=  \sum_l \sum_{s \in l, s \in S(t)} \E[t]{\mathbf{G}^s \beta^s \bone^s}  + \sum_l \sum_{s \in l, s \notin S(t)} \E[t]{\mathbf{G}^s \beta^s \bone^s}  + \beta^0\mathbf{1}_{n} \\   
        &=  \sum_l \sum_{s \in l, s \in S(t)} \beta^s\E[t]{\mathbf{G}^s \beta^s \bone^s}  + \sum_l \sum_{s \in l, s \notin S(t)} \beta^s\E[t]{\mathbf{G}^s \bone^s}  + \beta^0\mathbf{1}_{n} \\   
        &=  \sum_l \sum_{s \in l, s \in S(t)} \beta^s\mathbf{G}^s  + \sum_l \sum_{s \in l, s \notin S(t)} \beta^s\E[t]{\mathbf{G}^s \bone^s}  + \beta^0\mathbf{1}_{n} \\   
    \end{aligned}
    \label{proof:exp_expand}
\end{align}
\end{proof}

How we compute each element of $\E[t]{\mathbf{G}^s \bone^s}$ for $s \notin S(t)$.
\begin{align}
    \begin{aligned}
        \E[t]{G_i^s \bone^s} &= \frac{h_t^l}{L^l} \E[t]{\bone^s} \\
        &= \frac{h_t^l }{L^l} \cdot \frac{u^l}{n^l} L^l e^{-u^lL^l/n^l} \\ 
        &\approx \frac{h_t^l u^l}{n^l}\quad (\because\mbox{$u^l/n^l$ is small})  
    \end{aligned}
    \label{proof:exp_i}
\end{align}
Substituting \eq{\ref{proof:exp_expand}} to \eq{\ref{proof:exp_i}} gives the desired result.

\end{document}